\documentclass[prd,showpacs,twocolumn,preprintnumbers,amsmath,amssymb,superscriptaddress,floatfix,nofootinbib]{revtex4-2}%

\usepackage{graphicx}
\usepackage{bm}
\usepackage{amsmath}
\usepackage{amsfonts}
\usepackage{amssymb}
\usepackage{color}
\usepackage{multirow}
\usepackage{subfigure}
\usepackage{epstopdf}
\usepackage[colorlinks, citecolor=blue,anchorcolor=red,menucolor=red, linkcolor=red,filecolor=red,runcolor=red,urlcolor=blue,frenchlinks=red]{hyperref}

\begin{document}
	\title{Role of $\Xi(1690)$ in the $J/\psi\to\Xi^0\bar{\Lambda}K^0$ reaction}

	\author{Wen-Tao Lyu}\email{lvwentao9712@163.com}
	\affiliation{School of Physics, Zhengzhou University, Zhengzhou 450001, China}
	\affiliation{Departamento de Física Teórica and IFIC, Centro Mixto Universidad de Valencia-CSIC Institutos de Investigación de Paterna, 46071 Valencia, Spain}
	\vspace{0.5cm}
		
	\author{Lian-Rong Dai}\email{dailianrong@zjhu.edu.cn}
	\affiliation{School of Science, Huzhou Normal University, Huzhou, 313000, Zhejiang, China}
	\affiliation{Departamento de Física Teórica and IFIC, Centro Mixto Universidad de Valencia-CSIC Institutos de Investigación de Paterna, 46071 Valencia, Spain}\vspace{0.5cm}
	
	\author{Eulogio Oset}\email{oset@ific.uv.es}
	\affiliation{Departamento de Física Teórica and IFIC, Centro Mixto Universidad de Valencia-CSIC Institutos de Investigación de Paterna, 46071 Valencia, Spain}
	\vspace{0.5cm}
	
	

\begin{abstract}
Motivated by the recent BESIII measurements of the $J/\psi \to \Xi^0 \bar{\Lambda}K_S^0 + c.c.$ process, we investigate this reaction by considering the contributions from the $\Xi(1690)$, $\Lambda(1890)$ and $\Lambda(1830)$ resonances. The $\Xi(1690)$ state is dynamically generated from the $S$-wave pseudoscalar meson-octet baryon interactions within the chiral unitary approach. Our theoretical model provides a good description of the $\bar{\Lambda}K^0$, $\Xi^0 K^0$, and $\bar{\Lambda}\Xi^0$ invariant mass distributions. The results indicate that the $\Xi(1690)$ resonance, which was neglected in the experimental analysis by BESIII, plays a crucial role in this process. Furthermore, we evaluate the theoretical uncertainties of our model using the parametric bootstrap method. Future high-precision measurements of this process will further help to elucidate the properties of the $\Xi(1690)$, $\Lambda(1890)$ and $\Lambda(1830)$ states.

\end{abstract}
	
	\date{\today}
	
	\maketitle
	
\section{Introduction}\label{sec1}

While the ground-state octet and decuplet baryons have been well established, the spectrum of low-lying excited baryons with spin-parity $J^P=1/2^-$ remains puzzling. Aside from the well-known $N(1535)$ and $\Lambda(1405)$, the analogous low-lying $\Sigma(1/2^-)$ and $\Xi(1/2^-)$ states have not yet been firmly established~\cite{Wang:2024jyk}.

Currently, the Review of Particle Physics (RPP) lists two primary candidates for the $\Xi(1/2^-)$ state~\cite{ParticleDataGroup:2024cfk}. The first is the two-star $\Xi(1620)$, with a mass of approximately 1620~MeV and a width of $(32^{+8}_{-9})$~MeV. The second is the three-star $\Xi(1690)$, with a mass of $(1690\pm 10)$~MeV and a width of $(20\pm 15)$~MeV. However, their spin-parity quantum numbers lack definitive experimental confirmation. Historically, evidence for the $\Xi(1690)^0$ was first reported by the Belle and FOCUS Collaborations in 2002 via the resonant contribution in $\Lambda_c^+\to \Xi(1690)^0 K^+$~\cite{Belle:2001hyr,FOCUS:2005sye}. In 2005, the BaBar Collaboration observed the $\Xi(1690)^0$ in the $\Lambda_c^+\to \Lambda \bar{K}^0 K^+$ decay, measuring its mass and width as $(1684.7\pm1.3(\mathrm{stat})^{+2.2}_{-1.6}(\mathrm{syst}))$~MeV and $(8.1^{+3.9}_{-3.5}(\mathrm{stat})^{+1.0}_{-0.9}(\mathrm{syst}))$~MeV, respectively, and favoring a spin of $J=1/2$~\cite{BaBar:2006tck}. Subsequent BaBar analyses of $\Lambda_c^+\to \Xi^-\pi^+ K^+$ further supported the $J^P=1/2^-$ assignment for the $\Xi(1690)$~\cite{BaBar:2008myc}. More recently, in 2019, the Belle Collaboration reported the first observation of the doubly strange $\Xi(1620)^0$ baryon in the $\Xi_c^+\to\Xi^-\pi^+\pi^+$ process with a significance of $4\sigma$~\cite{Belle:2018lws}, and later identified both the $\Xi(1620)^0$ and $\Xi(1690)^0$ in the $\Xi\pi$ decay mode from the same process~\cite{Sumihama:2020mqa}.

On the theoretical front, the nature of the $\Xi(1/2^-)$ states has been extensively explored, though consensus remains elusive. Traditional constituent quark models typically predict the mass of the lowest $\Xi(1/2^-)$ state to be around 1800~MeV~\cite{Chao:1980em,Capstick:1986ter,Glozman:1995fu,Melde:2008yr}, which significantly overestimates the observed masses of both the $\Xi(1620)$ and $\Xi(1690)$. Alternatively, a large-$N_c$ QCD approach yields a mass of 1779~MeV~\cite{Schat:2001xr}, while the Skyrme model predicts two distinct $\Xi(1/2^-)$ states at 1616~MeV and 1658~MeV~\cite{Oh:2007cr}. Some studies classify the $\Xi(1690)$ as the first orbital excitation with $J^P=1/2^-$~\cite{Pervin:2007wa,Xiao:2013xi}. Conversely, chiral unitary approaches suggest a molecular picture; in Refs.~\cite{Li:2023olv,Ramos:2002xh,Miyahara:2016yyh,Garcia-Recio:2003ejq,Li:2025exm,Liu:2023jwo}, both the $\Xi(1620)$ and $\Xi(1690)$ emerge dynamically as $J^P=1/2^-$ resonances generated from the coupled-channel interactions of $\pi\Xi$, $\bar{K}\Lambda$, $\bar{K}\Sigma$, and $\eta\Xi$. 

Recently, the BESIII Collaboration analyzed the $J/\psi \to \Xi^0\bar{\Lambda}K_S^0 + c.c.$ reaction~\cite{BESIII:2025fuu} and determined its branching fraction to be $\mathcal{B} = (3.76 \pm 0.14 \pm 0.22) \times 10^{-5}$, where the uncertainties are statistical and systematic, respectively. However, the $\bar{\Lambda}K_S^0$ invariant mass distribution measured by BESIII exhibits a distinct structure around 1.67~GeV~(see Fig.~3 of Ref.~\cite{BESIII:2025fuu}), which is highly consistent with the mass of the predicted $\Xi(1690)$ state~\cite{Li:2023olv,Ramos:2002xh,Miyahara:2016yyh}.

In the present work, we demonstrate the significant role of the $\Xi(1690)$ in the $J/\psi \to \Xi^0\bar{\Lambda}K^0$ reaction, where the resonance is dynamically generated from the coupled-channel interactions of $\pi\Xi$, $\bar{K}\Lambda$, $\bar{K}\Sigma$, and $\eta\Xi$. Besides the $\Xi(1690)$ resonance in the $\bar{\Lambda}K^0$ channel, the experimental data also exhibits a prominent enhancement near the threshold of the $\Xi^0 K_S^0$ invariant mass spectrum. To properly describe these kinematic features, we further incorporate the contribution from the intermediate $\Lambda(1890)$ state. As a well-established baryon with spin-parity $J^P=3/2^+$, the $\Lambda(1890)$ is known to couple to the $\Xi K$ channel. This makes the decay $J/\psi \to \bar{\Lambda}\Lambda(1890) \to \bar{\Lambda}(\Xi^0 K^0)$ a natural mechanism to contribute to the observed $\Xi^0 K^0$ distribution. On the other hand, the $\Lambda(1830)$ state has a large branching fraction to $K\Xi$ ($\sim$56\%), and it is most likely to also contribute to the process. Hence, the contribution from the intermediate $\Lambda(1830)$ state with $J^P=5/2^-$ is also included in our calculation. By considering these contributions, we calculate the $\bar{\Lambda}K^0$, $\Xi^0K^0$, and $\bar{\Lambda}\Xi^0$ invariant mass distributions. Furthermore, we perform a rigorous analysis of the theoretical uncertainties using the parametric bootstrap method. 

This paper is organized as follows. In Sec.~\ref{sec2}, we present the theoretical formalism for the $J/\psi \to \Xi^0\bar{\Lambda}K^0$ decay. The numerical results and corresponding discussions are provided in Sec.~\ref{sec3}. Finally, a brief summary is given in Sec.~\ref{sec4}

\section{Formalism}\label{sec2}

\subsection{Role of $\Xi(1690)$ in $J/\psi\to\Xi^0\bar{\Lambda}K^0$}

In this section, we present the theoretical formalism. We begin with the strong decay process $J/\psi \to \Xi^0 \bar{\Lambda} K_S^0$, where the $K_S^0$ must strictly be a $K^0$ due to the nature of the strong decay. It can be seen that the $\bar{\Lambda} K^0$ pair can be produced via a $\bar{B}P$ interaction, where $\bar{B}$ is an anti-baryon belonging to the $1/2^-$ octet and $P$ is a pseudoscalar meson of the $0^-$ octet. To determine which $\bar{B}P$ states are produced in the $J/\psi \to \Xi^0 \bar{\Lambda} K^0$ reaction, we consider the $J/\psi$ to be an SU(3) singlet in the $u, d, s$ sector. By expressing the matrices for the $q\bar{q}$ in terms of mesons with the $\eta$-$\eta^\prime$ mixing of Ref.~\cite{Bramon:1992kr}, and the $qqq$ and $\bar{q}\bar{q}\bar{q}$ states for the baryon and anti-baryon octets, we have

\begin{eqnarray}\label{Eq:P_matrix}
	P=
	\left(
	\begin{array}{ccc}
		\frac{1}{\sqrt{2}}\pi^0 + \frac{1}{\sqrt{3}} \eta & \pi^+ & K^+ \\[2mm]
		\pi^- & -\frac{1}{\sqrt{2}} \pi^0 + \frac{1}{\sqrt{3}} \eta & K^0 \\[2mm]
		K^- & \bar{K}^0 & ~-\frac{1}{\sqrt{3}} \eta\\
	\end{array}
	\right), 
	\end{eqnarray}
	\begin{eqnarray}
	\label{Eq:B_matrix}
	B=
	\left(
	\begin{array}{ccc}
		\frac{1}{\sqrt{2}}\Sigma^0 + \frac{1}{\sqrt{6}} \Lambda & \Sigma^+ & p\\[2mm]
		\Sigma^- & -\frac{1}{\sqrt{2}} \Sigma^0 + \frac{1}{\sqrt{6}} \Lambda & n \\[2mm]
		\Xi^- & \Xi^0 & ~-\frac{2}{\sqrt{6}} \Lambda\\
	\end{array}
	\right) ,
\end{eqnarray}
\begin{eqnarray}
	\label{Eq:Bbar_matrix}
	\bar{B}=
	\left(
	\begin{array}{ccc}
		\frac{1}{\sqrt{2}}\bar{\Sigma}^0 + \frac{1}{\sqrt{6}} \bar{\Lambda} & \bar{\Sigma}^+ & \bar{\Xi}^+ \\[2mm]
		\bar{\Sigma}^- & -\frac{1}{\sqrt{2}} \bar{\Sigma}^0 + \frac{1}{\sqrt{6}} \bar{\Lambda }&  \bar{\Xi}^0\\[2mm]
		\bar{p} & \bar{n} & ~-\frac{2}{\sqrt{6}} \bar{\Lambda}\\
	\end{array}
	\right).
\end{eqnarray}

We analyze various SU(3) singlet structures formed by the trace of these products, specifically $\langle \bar{B} B P \rangle$, $\langle \bar{B} P B \rangle$, $\langle \bar{B} B\rangle \langle P \rangle$, $\langle \bar{B} P\rangle \langle B \rangle$, $\langle B P\rangle \langle \bar{B} \rangle$, and $\langle \bar{B} \rangle \langle B \rangle \langle P \rangle$. Among these, terms containing $\langle \bar{B} \rangle$ or $\langle B \rangle$ vanish because $B$ and $\bar{B}$ are traceless matrices. Furthermore, considering that the contribution from the $\langle \bar{B} B\rangle \langle P \rangle$ term is suppressed with respect to single trace structures, as shown using large-$N_c$ arguments~\cite{Manohar:1998xv,Abreu:2023yvf,Dai:2026zqn}, we focus on the $\langle BP\bar{B} \rangle$ and $\langle B  \bar{B}P \rangle$ structures in this study. 
This structure is equivalent to the one used in~\cite{He:2026mkf}
\begin{eqnarray}\label{Eq:LagJpsi}
    \mathcal{L}_\psi=\Tilde{D}  \left\langle \bar{B} \gamma_\mu \gamma_5 \{ P, B  \} \right\rangle \psi^\mu +\Tilde{F}  \left\langle \bar{B} \gamma_\mu \gamma_5 [ P, B  ] \right\rangle \psi^\mu.
\end{eqnarray}

It is worth noting that we must produce a $\Xi^0$ which means we specifically need the $\Xi^0 P\bar{B}$ and $\Xi^0 \bar{B}P$ structures. We find
\begin{eqnarray}\label{Eq:meson_baryon_A}
	\langle BP\bar{B} \rangle \Rightarrow \Xi^0\left\{\pi^-\bar{\Xi}^+-\frac{1}{\sqrt{2}}\pi^0\bar{\Xi}^0+\frac{1}{\sqrt{3}}\eta\bar{\Xi}^0-\frac{2}{\sqrt{6}}K^0\bar{\Lambda}\right\},\nonumber
\end{eqnarray}
and 
\begin{eqnarray}\label{Eq:meson_baryon_B}
	\langle B\bar{B}P \rangle \Rightarrow \Xi^0\left\{\bar{\Sigma}^-K^+-\frac{1}{\sqrt{2}}\bar{\Sigma}^0K^0+\frac{1}{\sqrt{6}}\bar{\Lambda}K^0-\frac{1}{\sqrt{3}}\bar{\Xi}^0\eta\right\}.\nonumber
\end{eqnarray}

To evaluate these two different structures, $\langle BP\bar{B} \rangle$ and $\langle B\bar{B}P \rangle$, we introduce distinct weights, $A$ and $B$, respectively. Thus, for practical purposes, we have

\begin{align}\label{Eq:jpsotoAB}
J/\psi \Rightarrow& \Xi^0\left\{A\left(\pi^-\bar{\Xi}^+-\frac{1}{\sqrt{2}}\pi^0\bar{\Xi}^0+\frac{1}{\sqrt{3}}\eta\bar{\Xi}^0-\frac{2}{\sqrt{6}}K^0\bar{\Lambda}\right) \right. \nonumber\\
&\left.+B\left(\bar{\Sigma}^-K^+-\frac{1}{\sqrt{2}}\bar{\Sigma}^0K^0+\frac{1}{\sqrt{6}}\bar{\Lambda}K^0-\frac{1}{\sqrt{3}}\bar{\Xi}^0\eta\right)\right\}.
\end{align}

Since $\Xi^0$ has isospin $I=1/2$ and $J/\psi$ $I=0$, the terms in the bracket in Eq.~(\ref{Eq:jpsotoAB}) will have $I=1/2$. Then, it is convenient to write the expression in bracket in isospin basis, and we take advantage to change the $B$ terms to the meson-baryon order when writing in isospin basis (change of sign for meson-baryon for $I=1/2$, with respect to baryon-meson). Using the isospin multiplets $(\bar{K}^0, -K^-)$, $(-\pi^+, \pi^0, \pi^-)$, $(\Xi^0, -\Xi^-)$, and $(-\Sigma^+, \Sigma^0, \Sigma^-)$, we obtain
\begin{equation}
	\begin{aligned}
		\left|\pi\Xi,I=\frac{1}{2},I_3=\frac{1}{2}\right>=\sqrt{\frac{2}{3}}\pi^+\Xi^--\sqrt{\frac{1}{3}}\pi^0\Xi^0,
	\end{aligned}
\end{equation}
\begin{equation}
	\begin{aligned}
		\left|\bar{K}\Sigma,I=\frac{1}{2},I_3=\frac{1}{2}\right>=-\left\{\sqrt{\frac{2}{3}}\Sigma^+K^--\sqrt{\frac{1}{3}}\Sigma^0\bar{K}^0\right\}.
	\end{aligned}
\end{equation}

Then the decay to the charge conjugate of Eq.~(\ref{Eq:jpsotoAB}) can be written as
\begin{align}\label{Eq:Hadronization_isospin}
	J/\psi \Rightarrow& \bar{\Xi}^0\left\{\left(\frac{-2A+B}{\sqrt{6}}\right)\bar{K}\Lambda^{I=1/2}+\left(\sqrt{\frac{3}{2}}A\right)\pi\Xi^{I=1/2} \right.\nonumber\\
	&\left.+\left(-\sqrt{\frac{3}{2}}B\right)\bar{K}\Sigma^{I=1/2}+\left(\frac{A-B}{\sqrt{3}}\right)\eta\Xi^{I=1/2}\right\}.
\end{align}

\begin{figure}[htbp]
	\centering
	
	\includegraphics[scale=0.45]{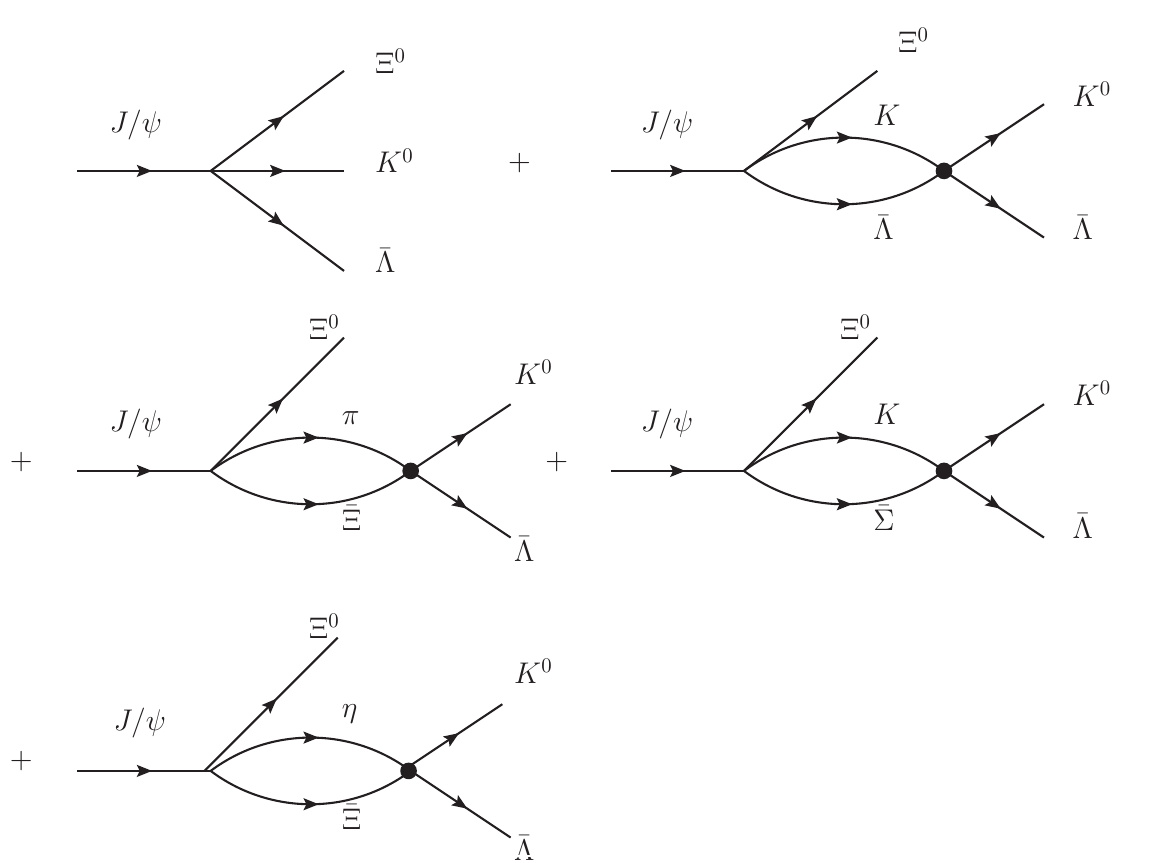}
	
	\caption{Mechanisms for tree level $J/\psi\to\Xi^0 \bar{\Lambda} K^0$ and rescattering of intermediate components.}\label{fig:jpsitoXiMB}
\end{figure}

The interaction of the coupled channels leading to the formation of the $\Xi(1690)$ state occurs in the $S$-wave; hence, we keep the $\bar{B}P$ pair in the $S$-wave. However, the $\Xi^0\bar{\Lambda}K^0$ final state has the spin-parity combination $J^P$ of $1/2^+$, $1/2^-$, and $0^-$, resulting in an overall positive parity. Therefore, a $P$-wave is required to match the negative parity of the $J/\psi$. This is automatically accounted for by using the spinor $v(p)$ for the production of the antiparticle and the structure of Eq.~(\ref{Eq:LagJpsi}).

We have then the transition matrix
\begin{eqnarray}\label{Eq:tJpsi}
    t=-\bar{u}_B(p)\gamma_\mu\gamma_5v_{\bar{\Xi}_0}(p^\prime)\epsilon^\mu\equiv \bar{u}_B(p) \gamma^i \gamma_5 v_{\bar{\Xi}^0}(p') \epsilon^i,
\end{eqnarray}
the last equation occurring in the $J/\psi$ rest frame, where $\epsilon^0 = 0$. We use the spinors up to order linear in momenta, 
\begin{equation}
u_B(p) = \begin{pmatrix}
	\chi_r  \\[8pt]
	\frac{\vec{\sigma}\cdot\vec{p}}{2m}\chi_r  
\end{pmatrix},~~~~~~v_{B^\prime}(p^\prime) = \begin{pmatrix}
	\frac{\vec{\sigma}\cdot\vec{p^\prime}}{2m^\prime}\chi_r^\prime  \\[8pt]
	\chi_r^\prime  
\end{pmatrix}.
\end{equation}

Based on this effective Lagrangian, we write the transition amplitude for the $J/\psi \to \bar{\Xi}^0\Lambda\bar{K}^0$ process as shown in Fig.~\ref{fig:jpsitoXiMB}~(now we have only spatial indices, considered all contravariant). The basic structure for the vertex of the tree level in Fig.~\ref{fig:jpsitoXiMB}~(first diagram) is
\begin{align}
\bar{u}(p) \gamma^i \gamma_5 v(p') \to 
\frac{P^i}{2m} + \frac{P'^i}{2m'} + i \epsilon_{ijk} \sigma_k \left( \frac{P'^j}{2m'} - \frac{P^j}{2m} \right).
\end{align}
All terms contribute for the tree level in Fig.~\ref{fig:jpsitoXiMB}, but in the rescattering diagrams, the terms with the momentum of the propagating baryon of the loop do not contribute, since the $t$ matrices are in $S$-wave. This said, we can write our total transition matrix for the sum of the mechanisms in Fig.~\ref{fig:jpsitoXiMB} as
\begin{equation}
	t_{\bar{\Xi}^0\bar{K}^0\Lambda}=\epsilon^i\tilde{t}^i_{\bar{\Xi}^0\bar{K}^0\Lambda},
\end{equation}
with
\begin{align}\label{Eq:ttilde_i}
	\tilde{t}^i_{\bar{\Xi}^0\bar{K}^0\Lambda}=&\frac{-2A+B}{\sqrt{6}}\frac{P^i_{\Lambda}}{2M_{\Lambda}}+H\frac{P^i_{\bar{\Xi}^0}}{2M_{\bar{\Xi}^0}} \nonumber\\
   &-i\frac{-2A+B}{\sqrt{6}}\epsilon_{ijk}\sigma_k\frac{P^j_{\Lambda}}{2M_{\Lambda}}+i\epsilon_{ijk}\sigma_kH\frac{P^j_{\bar{\Xi}^0}}{2M_{\bar{\Xi}^0}},
\end{align}
and
\begin{align}
	H=&\left\{\frac{-2A+B}{\sqrt{6}}+\frac{-2A+B}{\sqrt{6}}G_{\bar{K}\Lambda}t^{I=1/2}_{\bar{K}\Lambda,\bar{K}\Lambda}\right.\nonumber\\
	&\left.+\sqrt{\frac{3}{2}}AG_{\pi\Xi}t^{I=1/2}_{\pi\Xi,\bar{K}\Lambda}\right.\nonumber\\
	&\left.-\sqrt{\frac{3}{2}}BG_{\bar{K}\Sigma}t^{I=1/2}_{\bar{K}\Sigma,\bar{K}\Lambda}\right.\nonumber\\
	&\left. +\frac{A-B}{\sqrt{3}}G_{\eta\Xi}t^{I=1/2}_{\eta\Xi,\bar{K}\Lambda}\right\},
\end{align}
where we have already used the fact that $t_{\bar{M}\bar{B},\bar{M}\bar{B}}=t_{MB,MB}$ since we have the $MB$ amplitudes at hand. For the scattering amplitudes $t_{\bar{K}\Lambda, \bar{K}\Lambda}$, $t_{\pi\Xi, \bar{K}\Lambda}$, $t_{\bar{K}\Sigma, \bar{K}\Lambda}$, and $t_{\eta\Xi, \bar{K}\Lambda}$, we employ the chiral unitary approach with the coupled channels $\pi\Xi$ (1), $\bar{K}\Lambda$ (2), $\bar{K}\Sigma$ (3), and $\eta\Xi$ (4), as described in Refs.~\cite{Li:2023olv,Ramos:2002xh,Miyahara:2016yyh}. The transition potential is given by
\begin{align}\label{Eq:Vij}
	V_{ij} =& -C_{ij} \frac{1}{4f^2} (2\sqrt{s} - M_i - M_j)\nonumber\\
	&\times \left( \frac{M_i + E_i}{2M_i} \right)^{1/2} \left( \frac{M_j + E_j}{2M_j} \right)^{1/2},
\end{align}
with $f=93$~MeV, where the coefficients are given as,
\begin{equation}\label{Eq:Cij}
C_{ij} = \begin{pmatrix}
	2 & -\frac{3}{2} & -\frac{1}{2} & 0 \\[8pt]
	-\frac{3}{2} & 0 & 0 & -\frac{3}{2} \\[8pt]
	-\frac{1}{2} & 0 & 2 & \frac{3}{2} \\[8pt]
	0 & -\frac{3}{2} & \frac{3}{2} & 0
\end{pmatrix}.
\end{equation}
And the $T$ matrix is given by
\begin{equation}\label{Eq:BS}
T = [1-VG]^{-1}V,
\end{equation}
where $G_i$ is the standard meson-baryon loop function. We regularize it using the cut-off method with $q_{\text{max}} = 630$~MeV~\cite{Li:2023olv}, and the loop function can be written as~\cite{Guo:2005wp}
\begin{equation}
	\begin{aligned}
		G_i(s)=&\frac{2m_2}{16\pi^{2}s}\left\{\sigma\left(\arctan\frac{s+\Delta}{\sigma\lambda_{1}}+\arctan\frac{s-\Delta}{\sigma\lambda_{2}}\right)\right. \\
		&\left.-\left[\left(s+\Delta\right)\ln\frac{q_{\rm max}\left(1+\lambda_{1}\right)}{m_{1}}\right.\right. \\
		&\left.\left.+\left(s-\Delta\right)\ln\frac{q_{\rm max}\left(1+\lambda_{2}\right)}{m_{2}}\right]\right\},
	\end{aligned}
\end{equation}
where
\begin{equation}
	\sigma=\left[-\left(s-(m_{1}+m_{2})^{2}\right)\left(s-(m_{1}-m_{2})^{2}\right)\right]^{1/2},
\end{equation}
\begin{equation}
	\Delta=m_{1}^{2}-m_{2}^{2},
\end{equation}
\begin{equation}
	\lambda_1=\sqrt{1+\frac{m_{1}^{2}}{q_{\max}^{2}}},~~~~\lambda_2=\sqrt{1+\frac{m_{2}^{2}}{q_{\max}^{2}}},
\end{equation}
where $m_1$ and $m_2$ are the masses of meson and baryon, respectively.

\subsection{Role of intermediate state $\Lambda(1890)~(3/2^+)$}

\begin{figure}[htbp]
	\centering
	
	\includegraphics[scale=0.65]{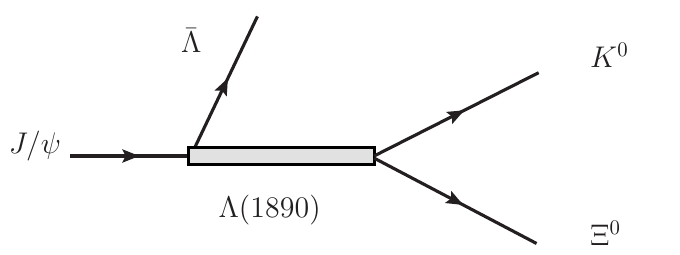}
	
	\caption{Mechanism for intermediate state $\Lambda(1890)$.}\label{fig:mechanism-Lambda1890}
\end{figure}

The excitation of the $\Lambda(1890)$ depicted in Fig.~\ref{fig:mechanism-Lambda1890} relies on a transition between a $1/2^+$ and a $3/2^+$ baryon state. Similar to the nucleon-$\Delta$ transition, this is achieved through the $S^\dagger$ transition spin operator, which is frequently employed in the $\Delta$-hole model for pion nuclear physics~\cite{Ericson:1988gk,Oset:1981ih}. It is defined as~\cite{Ericson:1988gk,Oset:1981ih,Duan:2024czu}
\begin{equation}
	\left\langle\frac{3}{2}M\left|S^\dagger_\nu\right|\frac{1}{2}m\right\rangle=C\left(\frac{1}{2},\mathbf{1},\frac{3}{2}; m,\nu,M\right)\left\langle\frac{3}{2}\left|\left|S^\dagger\right|\right|\frac{1}{2}\right\rangle,
\end{equation}
where $\nu$ is the spherical component of the rank-$\mathbf{1}$ operator $\vec{S}^\dagger$, which makes explicit use of the Wigner-Eckart theorem, and the reduced matrix element is defined as 1. For practical calculations, one uses
\begin{equation}\label{Eq:S}
	\sum_M S_l|M\rangle\langle M|S^\dagger_m=\frac{2}{3}\delta_{lm}-\frac{i}{3}\epsilon_{lms}\sigma_s,
\end{equation}
where $l$, $m$, and $s$ denote Cartesian components. The amplitude for the $\Lambda(1890)$ excitation, featuring a $P$-wave coupling at the $\Lambda(1890)\to K^0\Xi^0$ transition vertex, is given by~\cite{Lu:2016roh,Chen:2015sxa}
\begin{eqnarray}\label{Eq:t1890}
	t_{\Lambda(1890)}=&C \left\langle m^\prime\left|S_i\tilde{P}_{K^0i}\right|M\right\rangle\left\langle M\left|S_j^\dagger\epsilon_{j}\right|m\right\rangle D  \nonumber \\
	=&C \left\langle m^\prime\left|\frac{2}{3}\delta_{ij}-\frac{i}{3}\epsilon_{ijs}\sigma_s\right|m\right\rangle \tilde{P}_{K^0i}\epsilon_{j}D ,  
\end{eqnarray}
where 
\begin{equation}\label{Eq:D}
	D=\dfrac{1}{M_{\text{inv}}(K^0\Xi^0)-M_{\Lambda(1890)}+{i\Gamma_{\Lambda(1890)}}/{2}},
\end{equation}
and
\begin{equation}
	M_{\Lambda(1890)}=1870~\text{MeV},~~\Gamma_{\Lambda(1890)}=120.0~\text{MeV},
\end{equation}
where $\vec{\tilde{P}}_{K^0}$ is the momentum of the $K^0$ in the $K^0\Xi$ rest frame.

\subsection{Other possible intermediate states}

\begin{figure}[htbp]
	\centering
	
	\includegraphics[scale=0.65]{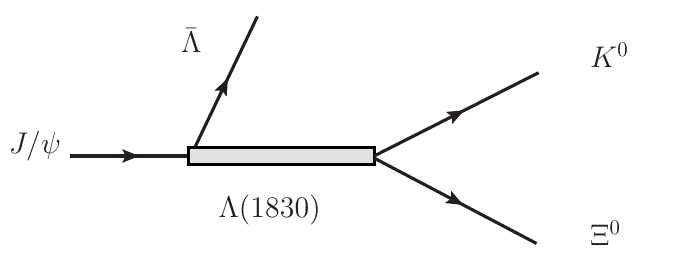}
	
	\caption{Mechanism for intermediate state $\Lambda(1830)$.}\label{fig:mechanism-Lambda1830}
\end{figure}

In the $\Xi^0 K^0$ invariant mass distribution, several other relevant resonances may also contribute, such as $\Lambda(1800)$, $\Lambda(1810)$, $\Lambda(1820)$, and $\Lambda(2000)$. However, no $K\Xi$ decay mode is reported for these states in the PDG. Moreover, among them, $\Lambda(1800)$ and $\Lambda(1810)$ are located at relatively low masses, whereas $\Lambda(2000)$ lies at a relatively high mass. Therefore, these states are not included in the present work. By contrast, the $\Lambda(1830)$ state can decay into $\Xi^0 K^0$, with a branching ratio of about $0.56$. Thus, we take this contribution into account. The corresponding mechanism for the $\Lambda(1830)~(5/2^-)$ contribution is depicted in Fig.~\ref{fig:mechanism-Lambda1830}. For reasons of spin and parity conservation, the coupling of $J/\psi$ to $\bar{\Lambda}\Lambda(1830)$ requires $L=3$. On the other hand, the coupling of $\Lambda(1830)$ to $K\Xi$ requires $L=2$. With such large angular momenta we do not expect interference with the other terms in the angular integrated mass distributions and we take an effective angle averaged amplitude
\begin{equation}\label{Eq:t1830}
	t_{\Lambda(1830)}=\frac{E}{M_\Lambda^4}\dfrac{\vec{P}_{\bar{\Lambda}}^3\vec{\tilde{P}}_{K^0}^2}{M_{\text{inv}}(K^0\Xi^0)-M_{\Lambda(1830)}+{i\Gamma_{\Lambda(1830)}}/{2}},  
\end{equation}
where $E$ is a free parameter and the $M_\Lambda^{-4}$ factor is included for dimensional reasons, such that $E$ has the same dimension as the $A$, $B$ and $C$ parameters. In our calculation, we take $M_{\Lambda(1830)}=1825$~MeV and $\Gamma_{\Lambda(1830)}=90$~MeV.

\subsection{Full decay amplitude}

We can now obtain the full $J/\psi \to \Xi^0\bar{\Lambda}K^0$ decay amplitude as follows
\begin{equation}
	t_{\text{Total}}=t_{\bar{\Xi}^0\bar{K}^0\Lambda}+t_{\Lambda(1890)}+\tilde{t}_{\Lambda(1830)}=\epsilon^i\tilde{t}^i+\tilde{t}_{\Lambda(1830)},
\end{equation}
where $\tilde{t}_{\Lambda(1830)}$ would contain all its angular dependence, with
\begin{align}\label{Eq:ttilde}
	\tilde{t}^i=&\frac{-2A+B}{\sqrt{6}}\frac{P_{\bar{\Lambda}}^i}{2M_{\bar{\Lambda}}}+\frac{2}{3}CD\tilde{P}^i_{K^0}+H\frac{P_{\Xi^0}^i}{2M_{\Xi^0}} \nonumber\\
 &-i\frac{-2A+B}{\sqrt{6}}\epsilon_{ijk}\sigma_k\frac{P_{\bar{\Lambda}}^j}{2M_{\bar{\Lambda}}} \nonumber\\
 &+\frac{i}{3}CD\epsilon_{ijk}\sigma_k\tilde{P}^j_{K^0} \nonumber\\
 &+iH\epsilon_{ijk}\sigma_k\frac{P_{\Xi^0}^j}{2M_{\Xi^0}},
\end{align}
where with respect to Eq.~(\ref{Eq:ttilde_i}), meant for the $J/\psi \to \bar{\Xi}^0 \Lambda \bar{K}^0$, we have changed $\vec{P}_\Lambda$ and $\vec{P}_{\bar{\Xi}^0}$ to $\vec{P}_{\bar{\Lambda}}$ and $\vec{P}_{\Xi^0}$ of the observed $J/\psi \to \Xi^0 \bar{\Lambda} K^0$ reaction.

By summing and averaging $|t_{\text{Total}}|^2$ over the particle spins, considering that the $t_{MB}$ amplitudes are spin-independent, and that we do not have interference of $\tilde{t}_{\Lambda(1830)}$ with the other terms, we obtain
\begin{align}\label{Eq:ttotal}
&\overline{\sum}\sum|t_{\text{Total}}|^2
=\frac{1}{3}\sum_{\text{spin}}\tilde{t}^i\tilde{t}^{*i}+|t_{\Lambda(1830)}|^2 \nonumber\\
&=\frac{2}{3}\left\{\left(\frac{-2A+B}{\sqrt{6}}\right)^2\frac{\vec{P}_{\bar{\Lambda}}^2}{4M_{\bar{\Lambda}}^2}+\frac{4}{9}C^2|D|^2\vec{\tilde{P}}_{K^0}^2+|H|^2\frac{\vec{P}_{\Xi^0}^2}{4M_{\Xi^0}^2}\right. \nonumber\\
&~~~+\left(\frac{-2A+B}{\sqrt{6}}\right)\frac{2}{3}C~2~\textbf{Re}(D)\frac{\vec{P}_{\bar{\Lambda}}\cdot\vec{\tilde{P}}_{K^0}}{2M_{\bar{\Lambda}}}\nonumber\\
&~~~+\left(\frac{-2A+B}{\sqrt{6}}\right)~2~\textbf{Re}(H)\frac{\vec{P}_{\bar{\Lambda}}\cdot\vec{P}_{\Xi^0}}{2M_{\bar{\Lambda}}2M_{\Xi^0}} \nonumber\\
&\left.~~~+\frac{2}{3}C~2~\textbf{Re}(DH^*)\frac{\vec{P}_{\Xi^0}\cdot\vec{\tilde{P}}_{K^0}}{2M_{\Xi^0}}\right\}\nonumber\\
&+\frac{4}{3}\left\{\left(\frac{-2A+B}{\sqrt{6}}\right)^2\frac{\vec{P}_{\bar{\Lambda}}^2}{4M_{\bar{\Lambda}}^2}+\frac{1}{9}C^2|D|^2\vec{\tilde{P}}_{K^0}^2+|H|^2\frac{\vec{P}_{\Xi^0}^2}{4M_{\Xi^0}^2}\right. \nonumber\\
&~~~-\frac{1}{3}\left(\frac{-2A+B}{\sqrt{6}}\right)C~2~\textbf{Re}(D)\frac{\vec{P}_{\bar{\Lambda}}\cdot\vec{\tilde{P}}_{K^0}}{2M_{\bar{\Lambda}}}\nonumber\\
&~~~-\left(\frac{-2A+B}{\sqrt{6}}\right)~2~\textbf{Re}(H)\frac{\vec{P}_{\bar{\Lambda}}\cdot\vec{P}_{\Xi^0}}{2M_{\bar{\Lambda}}2M_{\Xi^0}} \nonumber\\
&\left.~~~+\frac{1}{3}C~2~\textbf{Re}(DH^*)\frac{\vec{P}_{\Xi^0}\cdot\vec{\tilde{P}}_{K^0}}{2M_{\Xi^0}}\right\}+|t_{\Lambda(1830)}|^2,
\end{align}
where $\vec{P}_{\Xi^0}$ and $\vec{P}_{\bar{\Lambda}}$ are the momenta of the $\Xi^0$ and $\bar{\Lambda}$ in the $J/\psi$ rest frame, respectively, and $\vec{\tilde{P}}_{K^0}$ is the momentum of the $K^0$ in the $K^0\Xi^0$ rest frame, given by
\begin{equation}
\left|\vec{P}_{\Xi^0}\right|=\dfrac{\lambda^{1/2}(m_{J/\psi}^2,M_{\Xi^0}^2,M^2_{\text{inv}}(K^0\bar{\Lambda}))}{2m_{J/\psi}},
\end{equation}
\begin{equation}
\left|\vec{P}_{\bar{\Lambda}}\right|=\dfrac{\lambda^{1/2}(m_{J/\psi}^2,M_{\bar{\Lambda}}^2,M^2_{\text{inv}}(K^0\Xi^0))}{2m_{J/\psi}},
\end{equation}
\begin{equation}
\left|\vec{\tilde{P}}_{K^0}\right|=\dfrac{\lambda^{1/2}(M^2_{\text{inv}}(K^0\Xi^0),m_{K^0}^2,M^2_{\Xi^0})}{2M_{\text{inv}}(K^0\Xi^0)},
\end{equation}
\begin{align}
	\vec{P}_{\bar{\Lambda}}\cdot\vec{P}_{\Xi^0}=\frac{1}{2}\left\{M_{\bar{\Lambda}}^2+M_{\Xi^0}^2+2E_{\Xi^0}E_{\bar{\Lambda}}-M^2_{\text{inv}}(\Xi^0\bar{\Lambda})\right\}.
\end{align}
Since $\vec{P}_{\Xi^0}$ and $\vec{\tilde{P}}_{K^0}$ are evaluated in different reference frames, we must apply a Lorentz boost. This yields~\cite{Crdoba1995ProjectileDE}
\begin{equation}
	\vec{P}_{\Xi^0}=\left[\left(\frac{\tilde{E}_{J/\psi}}{m_{J/\psi}}-1\right)\frac{\vec{\tilde{P}}_{\Xi^0}\cdot\vec{\tilde{P}}_{J/\psi}}{\vec{\tilde{P}}_{J/\psi}^2}-\frac{\tilde{E}_{\Xi^0}}{m_{J/\psi}}\right]\vec{\tilde{P}}_{J/\psi}+\vec{\tilde{P}}_{\Xi^0},
\end{equation}
and where the tilde refers to the $K^0\Xi^0$ rest frame,
\begin{equation}
	\vec{\tilde{P}}_{K^0}=-\vec{\tilde{P}}_{\Xi^0}.
\end{equation}
Then, we have
\begin{align}
\vec{\tilde{P}}_{K^0}\cdot\vec{P}_{\Xi^0}=&-\left[\left(\frac{\tilde{E}_{J/\psi}}{m_{J/\psi}}-1\right)\frac{\vec{\tilde{P}}_{\Xi^0}\cdot\vec{\tilde{P}}_{J/\psi}}{\vec{\tilde{P}}_{J/\psi}^2} \right.\nonumber\\
&\left.-\frac{\tilde{E}_{\Xi^0}}{m_{J/\psi}}\right]\vec{\tilde{P}}_{\Xi^0}\cdot\vec{\tilde{P}}_{J/\psi}-\left|\vec{\tilde{P}}_{\Xi^0}\right|^2,
\end{align}
where
\begin{align}
	\vec{\tilde{P}}_{\Xi^0}\cdot\vec{\tilde{P}}_{J/\psi}=\frac{1}{2}\left\{M_{\Xi^0}^2+M_{\bar{\Lambda}}^2+2\tilde{E}_{\Xi^0}\tilde{E}_{\bar{\Lambda}}-M^2_{\text{inv}}(\Xi^0\bar{\Lambda})\right\},
\end{align}
and
\begin{align}
	\vec{\tilde{P}}_{J/\psi}=\vec{\tilde{P}}_{\bar{\Lambda}},
\end{align}
with
\begin{equation}
	\left|\vec{\tilde{P}}_{\bar{\Lambda}}\right|=\dfrac{\lambda^{1/2}(m^2_{J/\psi},M_{\bar{\Lambda}}^2,M^2_{\text{inv}}(K^0\Xi^0))}{2M_{\text{inv}}(K^0\Xi^0)}.
\end{equation}

Similarly, for the $\bar{\Lambda}$ momentum, we obtain
\begin{equation}
	\vec{P}_{\bar{\Lambda}}=\left[\left(\frac{\tilde{E}_{J/\psi}}{m_{J/\psi}}-1\right)\frac{\vec{\tilde{P}}_{\bar{\Lambda}}\cdot\vec{\tilde{P}}_{J/\psi}}{\vec{\tilde{P}}_{J/\psi}^2}-\frac{\tilde{E}_{\bar{\Lambda}}}{m_{J/\psi}}\right]\vec{\tilde{P}}_{J/\psi}+\vec{\tilde{P}}_{\bar{\Lambda}},
\end{equation}
which leads to
\begin{align}
\vec{\tilde{P}}_{K^0}\cdot\vec{P}_{\bar{\Lambda}}=\left(\frac{\tilde{E}_{J/\psi}}{m_{J/\psi}}-\frac{\tilde{E}_{\bar{\Lambda}}}{m_{J/\psi}}\right)\vec{\tilde{P}}_{\bar{\Lambda}}\cdot\vec{\tilde{P}}_{K^0},
\end{align}
with
\begin{align}
	\vec{\tilde{P}}_{\bar{\Lambda}}\cdot\vec{\tilde{P}}_{K^0}=\frac{1}{2}\left\{M_{\bar{\Lambda}}^2+M_{K^0}^2+2\tilde{E}_{K^0}\tilde{E}_{\bar{\Lambda}}-M^2_{\text{inv}}(K^0\bar{\Lambda})\right\}.
\end{align}

According to the notation 1 to $\Xi^0$, 2 to $K^0$ and 3 to $\bar{\Lambda}$ and applying the standard kinematics formulas from the RPP~\cite{ParticleDataGroup:2024cfk}, we have
\begin{equation}
	\dfrac{d^2\Gamma}{dM_{12}dM_{23}} = \frac{1}{(2\pi)^3}\dfrac{2M_{\bar{\Lambda}} 2M_{\Xi^0}}{32m_{J/\psi}^3}\overline{\sum}\sum|t_{\text{Total}}|^2~2M_{12}~2M_{23},
\end{equation}
using the Mandl and Shaw normalization for the meson and baryon fields~\cite{Mandl:1985bg}.

We can obtain $d\Gamma/dM_{12}$ by integrating $d^2\Gamma/(dM_{12}dM_{23})$ over $M_{23}$ within the kinematic limits specified in the RPP~\cite{ParticleDataGroup:2024cfk}. Permuting the indices allows us to evaluate all three invariant mass distributions. We use $M_{12}$ and $M_{23}$ as independent variables, and determine $M_{13}$ using the kinematic relation $M_{12}^2+M_{13}^2+M_{23}^2=m_{J/\psi}^2+M_{\Xi^0}^2+m_{K^0}^2+M_{\bar{\Lambda}}^2$ to get $M_{13}$ from them.

\section{Results}\label{sec3}

\begin{figure}
	\subfigure[]{
			\centering
				\includegraphics[scale=0.65]{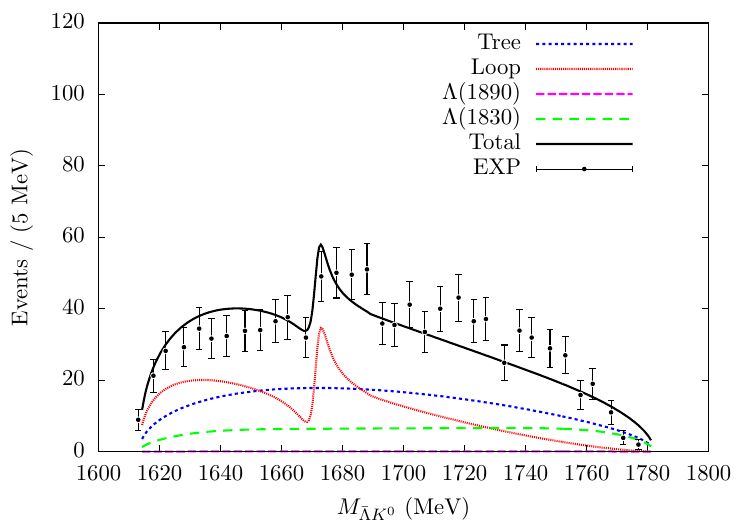}
			}
	\subfigure[]{
			\centering
				\includegraphics[scale=0.6]{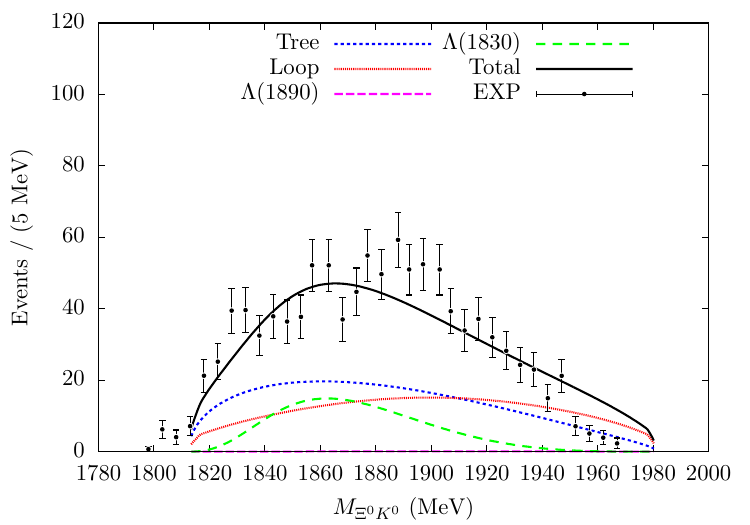}
			}
	\subfigure[]{
			\centering
			    \includegraphics[scale=0.6]{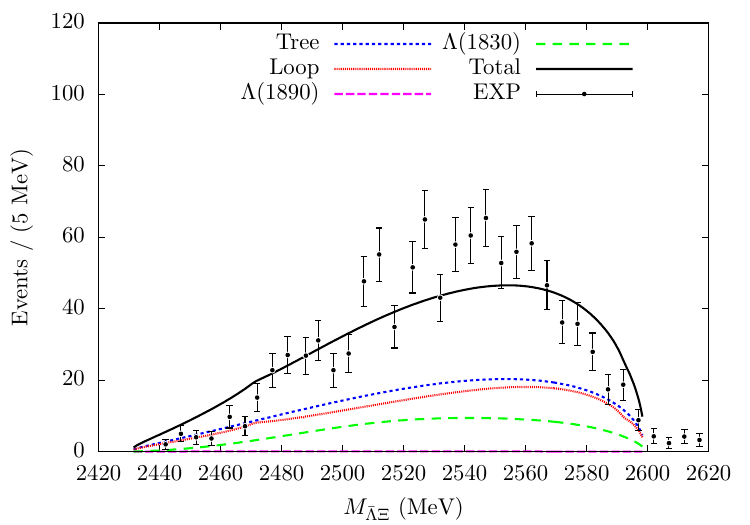}
			}
	\caption{Invariant mass distributions of $\bar{\Lambda}K^0$ (a), $\Xi^0K^0$ (b), and $\bar{\Lambda}\Xi^0$ (c) compared with the BESIII experimental data~\cite{BESIII:2025fuu} from which the background contribution has been subtracted.}\label{fig:dwidth}
\end{figure}

\begin{figure}
	\subfigure[]{
		\centering
		\includegraphics[scale=0.6]{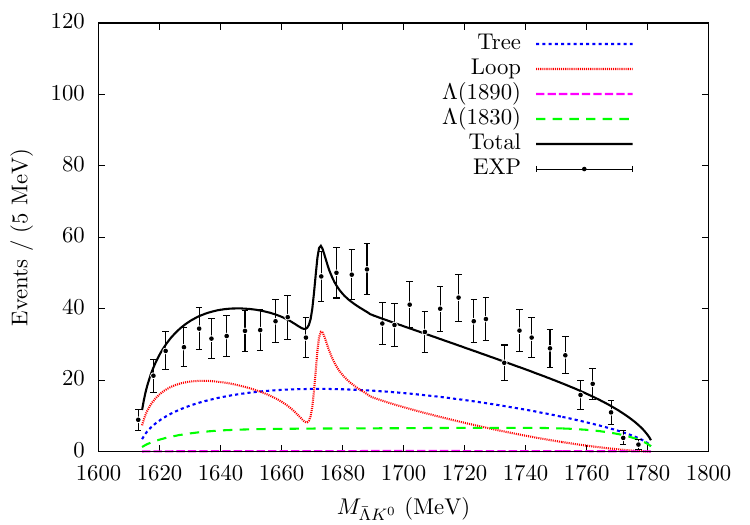}
	}
	\subfigure[]{
		\centering
		\includegraphics[scale=0.6]{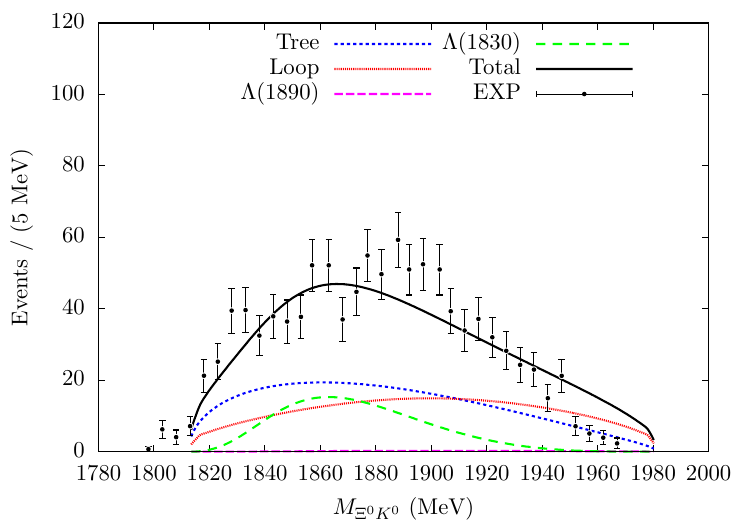}
	}
	\subfigure[]{
		\centering
		\includegraphics[scale=0.6]{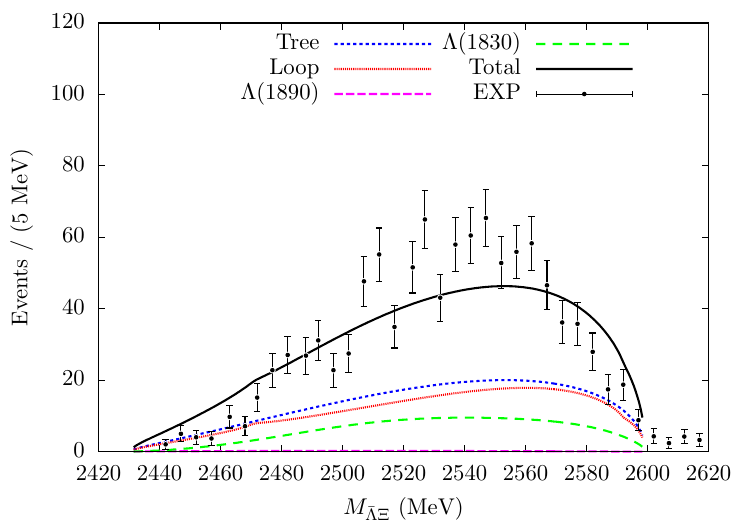}
	}
	\caption{Invariant mass distributions of $\bar{\Lambda}K^0$ (a), $\Xi^0K^0$ (b), and $\bar{\Lambda}\Xi^0$ (c) compared with the BESIII experimental data~\cite{BESIII:2025fuu} from which the background contribution has been subtracted, obtained from the 5-parameter fit which includes the interference phase $\phi$.}\label{fig:dwidth_phase}
\end{figure}

\begin{figure}
	\subfigure[]{
		\centering
		\includegraphics[scale=0.6]{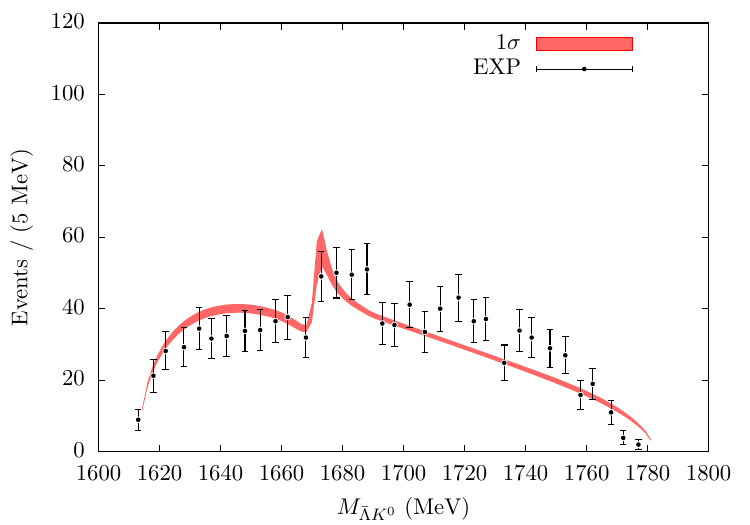}
	}
	\subfigure[]{
		\centering
		\includegraphics[scale=0.6]{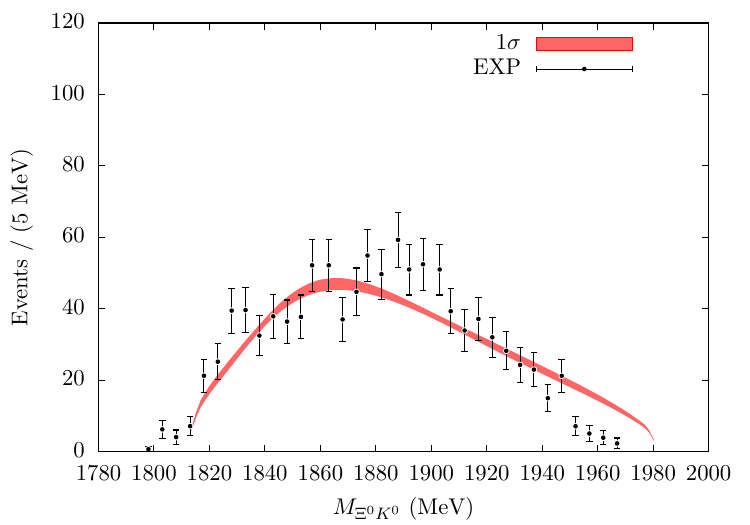}
	}
	\subfigure[]{
		\centering
		\includegraphics[scale=0.6]{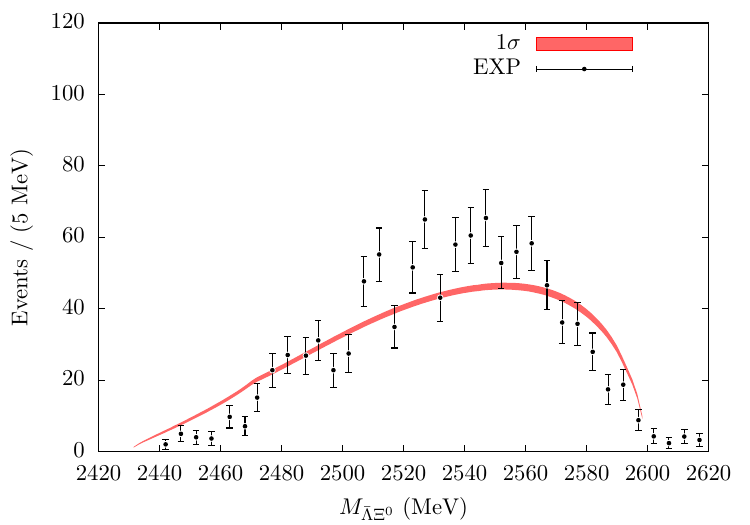}
	}
	\caption{Theoretical uncertainties for the invariant mass distributions of $\bar{\Lambda}K^0$ (a), $\Xi^0K^0$ (b), and $\bar{\Lambda}\Xi^0$ (c) evaluated using the parametric bootstrap method.}\label{fig:dwidth_error}
\end{figure}

In our model, there are four free parameters: $A$, $B$, $C$ and $E$. Fitting these parameters to the experimental data\footnote{Given the fact that in the BESIII experiment some points are given outside the allowed phase space, we have removed for the fit the points within 15 MeV in the extremes of the distributions.} yields $A = 2.40~\text{ MeV}^{-1}$, $B = 0.34~\text{ MeV}^{-1}$, $C = 0.02~\text{ MeV}^{-1}$ and $E = -33.34~\text{ MeV}^{-1}$, giving dimension MeV to the events number, as if it were a width, resulting in a $\chi^2/\text{d.o.f.} = 196.95/(89-4) = 2.32$.

In Fig.~\ref{fig:dwidth}, we present the invariant mass distributions for $\bar{\Lambda}K^0$, $\Xi^0K^0$, and $\bar{\Lambda}\Xi^0$. The blue dotted curves represent the tree-level contributions, while the red dotted curves indicate the contributions from the meson-baryon interaction associated with the $\Xi(1690)$ resonance. Notably, a distinct structure emerges around $1.67 \text{ GeV}$ in the $\bar{\Lambda}K^0$ invariant mass distribution, which is due to the dynamically generated $\Xi(1690)$ state. The magenta dashed curves display the contribution from the intermediate $\Lambda(1890) (3/2^+)$ state and the green dashed curves show the contribution from the intermediate $\Lambda(1830) (5/2^-)$. A clear peak can be observed in the $\Xi^0K^0$ invariant mass distribution, which is associated with the $\Lambda(1830)$ state. As we can see, the contribution of the $\Lambda(1890)$ state is negligible. This would be compatible with the small branching ratio of this resonance to $K\Xi$, of the order of 1\%. The black solid curves represent the total amplitude. A broad peak is also visible in the $\Xi^0K^0$ invariant mass distribution. We would like to stress that the singular structure in the region of the $\Xi(1690)$, with a dip followed by a peak, is also present in the experiment. This manifestation of the resonance is similar to the one observed in the $\Xi_c^+\to\pi^+\pi^+\Xi^-$~\cite{Belle:2018lws}, which was also reproduced theoretically in Ref.~\cite{Li:2023olv}. The shape of the resonance could be different in other reactions, as shown in the study of the $\Xi_c^+\to\Lambda\bar{K}^0\pi^+$ in Ref.~\cite{Li:2025exm}.

To eventually improve the fit, we take advantage of the phase freedom in the $C$ term in Eqs.~(\ref{Eq:t1890}), (\ref{Eq:ttilde}) and (\ref{Eq:ttotal}) by allowing $C$ to acquire a complex phase
\begin{equation}
	C\rightarrow C e^{i\phi}.
\end{equation}

With this modification, the model now has five free parameters. The updated fit yields $A = 2.39~\text{MeV}^{-1}$, $B = 0.34~\text{MeV}^{-1}$, $C = 0.03~\text{MeV}^{-1}$, $E = -32.02~\text{MeV}^{-1}$, $\phi = 0.24\pi$, with new $\chi^2/\text{d.o.f.} = 196.84/(89-5) = 2.34$. The corresponding results are displayed in Fig.~\ref{fig:dwidth_phase}. As we can see, including this extra phase still makes the contribution of the $\Lambda(1890)$ negligible.

To estimate the theoretical uncertainties and evaluate the model's sensitivity to the experimental data, we employ a parametric bootstrap (resampling) approach~\cite{Albaladejo:2016hae,Efron:1986hys,Press:1992}. Specifically, we generate $1000$ sets of pseudo-data by generating new centroids of the data with Gaussian distributions together with the original experimental errors. Our model is then refitted to each of these pseudo-data sets to obtain new sets of parameters and from them the new distributions. These results are presented in Fig.~\ref{fig:dwidth_error}, where the bands represent the $1\sigma$ confidence level.

\section{ Conclusions }\label{sec4}

Recently, the BESIII Collaboration reported the measurements of the $J/\psi \to \Xi^0\bar{\Lambda}K_S^0 + c.c.$ reaction~\cite{BESIII:2025fuu}. Notably, the measured $\bar{\Lambda}K_S^0$ invariant mass distribution reveals a clear structure around 1.67~GeV, which aligns well with the predicted mass of the $\Xi(1690)$ resonance.

Motivated by this observation, we have performed a theoretical study of the $J/\psi \to \Xi^0\bar{\Lambda}K^0$ process. We specifically focus on the contribution of the $\Xi(1690)$, which is dynamically generated from the $S$-wave coupled-channel interactions of $\pi\Xi$, $\bar{K}\Lambda$, $\bar{K}\Sigma$, and $\eta\Xi$ within the chiral unitary approach. To evaluate the production of the $\Xi(1690)$, we treat the $J/\psi$ as an SU(3) singlet and identify the dominant mechanisms in which it decays into an SU(3) octet baryon and a pseudoscalar-baryon pair. Alongside the $\Xi(1690)$, the contributions from the intermediate $\Lambda(1890)$ and $\Lambda(1830)$ states are also taken into account.

Our results yield a satisfactory description of the $\bar{\Lambda}K^0$, $\Xi^0K^0$, and $\bar{\Lambda}\Xi^0$ invariant mass distributions simultaneously. This agreement not only supports the molecular nature of the $\Xi(1690)$ but also underscores its indispensable role in this decay process. Given that the properties of the $\Xi(1690)$ are not yet firmly established and the current BESIII data are subject to large statistical uncertainties, future high-precision measurements at facilities like Belle II and the proposed Super Tau-Charm Facility (STCF) will be most welcome. Such experiments will shed further light on the underlying reaction mechanisms and the precise nature of the $\Xi(1690)$.

Note added in proof: The formalism used in this paper was inspired 
by the work of Ikeno~\cite{Ikeno:2026vfs}, which, for various reasons, was delayed 
before being made publicly available online.

\section*{Acknowledgments}

We would like to acknowledge useful discussions with Profs. Natsumi Ikeno, En Wang and De-Min Li. This work was supported by the National Key R\&D Program of China (Grant No. 2024YFE0105200), the Natural Science Foundation of Henan (Grant No. 252300423951), the National Natural Science Foundation of China (Grant No. 12475086 and No. 12575082), and the Zhengzhou University Young Student Basic Research Projects for PhD students (Grant No. ZDBJ202522). Wen-Tao Lyu acknowledges the support of the China Scholarship Council. This work is also partly supported by the Spanish Ministerio de Economia y Competitividad~(MINECO) and European FEDER funds under Contracts No. FIS2017-84038-C2-1-PB, PID2020-112777GB-I00, and by Generalitat Valenciana under contract PROMETEO/2020/023. This project has received funding from the European Union Horizon 2020 research and innovation program under the program H2020-INFRAIA-2018-1, grant agreement No. 824093 of the STRONG-2020 project.

\end{document}